\documentclass{ws-ijbc}
\usepackage{graphicx}
\usepackage{epstopdf}
\usepackage[mathscr]{euscript}
\begin{document}
\catchline{}{}{}{}{} 

\markboth{Author's Name}{Paper Title}
\title{Period doubling, information entropy, and estimates for Feigenbaum's constants}

\author{Reginald D. Smith}

\address{Citizen Scientists League\\
rsmith@citizenscientistsleague.com\footnote{PO Box 10051, Rochester, NY 14610}}

\maketitle

\begin{history}
\received{(to be inserted by publisher)}
\end{history}

\begin{abstract}
The relationship between period doubling bifurcations and Feigenbaum's constants has been studied for nearly 40 years and this relationship has helped uncover many fundamental aspects of universal scaling across multiple nonlinear dynamical systems. This paper will combine information entropy with symbolic dynamics to demonstrate how period doubling can be defined using these tools alone. In addition, the technique allows us to uncover some unexpected, simple estimates for Feigenbaum's constants which relate them to $\log 2$ and the golden ratio, $\varphi$, as well as to each other.
\end{abstract}


\maketitle

\section{Period doubling and Feigenbaum's Constants}

The discovery of Feigenbaum's constants, where Mitchell Feigenbaum demonstrated that unique, universal constants are linked to successive measurements of iterable maps of quadratic functions, was a groundbreaking discovery in the study of nonlinear dynamics. By studying iterable mappings of difference equations of the form

\begin{equation}
x_{n+1}=\mu f(-x_n^2)
\end{equation}

most popularly, the logistic population growth map,

\begin{equation}
x_{n+1}=\mu x_n(1-x_n)
\end{equation}

he was able to demonstrate the existence of two key constants \citep{feigenbaum0, feigenbaum1}. The first, sometimes called the first constant, $\delta$, is defined as the limit of the ratios of the parameter intervals between bifurcation points

\begin{equation}
\delta = \lim_{N\rightarrow \infty} \frac{\mu_{N+1}-\mu_{N}}{\mu_{N+2}-\mu_{N+1}} = 4.669202\dots 
\label{feigenbaumfirst}
\end{equation}

The second, sometimes called the second constant, $\alpha$, is defined as the ratio of the maximum distance between points in each $2^n$ periodic cycle. This distance reduced in a near constant ratio after each bifurcation to where $\alpha=2.50291$\dots. In \citep{feigenbaum1} it was recognized this does not only apply to functions of the form $x_{n+1}=\mu f(-x_n^2)$ but to all functions with quadratic maximums such as the sine map $x_{n+1}=\mu \sin(x_n)$ or entropy map $x_{n+1}=-\mu(x_n\log x_n  + (1-x_n)\log (1-x_n))$.

The foundations of Feigenbaum's work were solidified in \citep{collet1,lanford,feigenbaum2} and the universality of the constants was generally recognized. In addition, for cubic and other types of iterable maps, different versions of Feigenbaum's constants were found to exist. Feigenbaum's constants, however, have not received a more exact definition than the current numerical calculation. Whether they are transcendental or can be fully expressed in terms of other constants or expressions remains unknown.

\section{Period doubling and conditional entropy}

An alternate way to understand and investigate period doubling relies on the use of information entropy. In particular, one can define a period doubling cascade only using the various order entropies and a symbolic dynamical system with a binary alphabet.

Symbolic dynamics is a common way to represent sequences. Typically in symbolic dynamics, we study a bi-infinite sequence of symbols, all of which belong to a set known as the alphabet. For a binary sequence, the alphabet is $\mathscr{A}=\{0,1\}$. The dynamics of such a sequence are carried out by the use of a shift map, $\sigma$, which when applied to a position on a sequence ($\sigma(x)$) effectively maps the sequence onto a new sequence where the position becomes the next symbol in first sequence ($y=\sigma(x)$).

From this methodology a periodic orbit, or limit cycle, of period $T$ is defined as $\sigma^T(x)=x$. So a limit cycle $T=2^n$ is represented by $\sigma^{2^n}(x)=x$.

The information entropy of a bi-infinite sequence with a finite alphabet, $\mathscr{A}$ can be calculated in the regular fashion. In particular, the Shannon entropy of the sequence can be defined by

\begin{equation}
H_1=-\sum_{k=1}^M p_k \log p_k
\end{equation}

Where $M$ is the number of elements in $\mathscr{A}$. This definition of entropy can be extended to $N$-grams, $N$ symbols in length, to define the $N$-order entropy. In the binary sequence case this is

\begin{equation}
H_N=-\sum_{k=1}^{2^N} p_k \log p_k
\end{equation}

Finally, we can define the conditional entropy of order $N$, $H(N)$, as the difference between entropies in two consecutive orders

\begin{equation}
H(N) = H_N - H_{N-1}
\end{equation}

By definition, the Shannon entropy and $H(1)$ are identical: $H(1)=H_1$. As $N$ increases, $H(N)$ must stay constant or monotonically decrease. It is commonly known that the number of unique binary sequences of length $L$, $W_L$, can be given by $2^L$. In addition, if one knows $H_1$, the number of unique sequences can be further narrowed to \citep{shannonweaver}

\begin{equation}
W_L=2^{LH_1}
\end{equation}

This result, however, assumes no correlations between the appearance of symbols given previous symbols. Further estimates can be shown using the conditional entropies from $H(2)$ \citep{kolmogorov} to $H(N)$  where $N \leq L$ \citep{smith}

\begin{equation}
W_L = 2^{LH(N)}
\end{equation} 

In the case where $LH(N)=0$, there is only one possible sequence and $W_L=1$.

\subsection{Period doubling limit cycles}

For the purposes of this paper, all entropies will be defined assuming sequences with a binary alphabet $\mathscr{A}=\{0,1\}$. For completely random sequences, $H_1=\log 2$, $H_2=2\log 2$, $H_3=3\log 2$, etc. Conditional entropies for all $N$ are equal to $\log 2$ under these circumstances. 

The investigation of cycles of $T=2^{N-1}$ begins when you fulfill two conditions: first, you set the entropy of all orders less than $N$ to the values they would have assuming a random sequence. This would be the order of entropy times $\log 2$. Second, you equate the entropy of order $N$ with the entropy of order $N-1$. For example, $H_2=H_1$. Under the definition of conditional entropy, $H(N)=0$, for all conditional entropies of order $N$ and higher and there is a single solution for sequences with length at least $2^{N-1}$.

For the case, $H_2=H_1$, we find that we have a period two limit cycle ``010101''. This is due to $H_2$ having to have only two, equally probable length two symbols that must fulfill the conditions of equal numbers of ``0'' and ``1'' as dictated by $H_1=\log 2$. The same can be seen where $H_1=\log 2$, $H_2=2\log 2$ and $H_3=H_2$. Here, where $N=3$ we have a period four limit cycle represented by infinite repeats of ``1001''. This can continue for consecutive orders of entropy. 

As an interesting side note, in each of these limit cycles, due to the constraints of randomness imposed on the lower orders (less than $N$) of entropy, each possible permutation of binary words from length 1 to $N-1$ appears in the sequence of length $2^{N-1}$ with a probability $1/2^M$, where $M$ is the length of the word. Each possible $N-1$ length permutation thus appears once and therefore, the number of possible permutations of a limit cycle of period $2^{N-1}$ can be given by the number of possible DeBruijn sequences
\begin{equation}
\frac{2^{2^{N-2}}}{2^{N-1}}.
\end{equation}

This is also important since the digraph of transition states for words of length $N-1$ for the DeBruijn sequence can be used as the shift digraph for the symbolic dynamics. This digraph is regular with of degree 2 and thus has an index eigenvalue, $\lambda$, equal to 2 \citep{spectral}. This allows us to define the topological entropy of all sequences in the period doubling cascade to be $H_T(x) = \log \lambda = \log 2$.

\section{Iterative maps and period doubling}

The investigation of the period doubling bifurcation can be combined with the entropies defined previously. In particular, for each limit cycle, the entropies are fixed within the interval between the parameter values that define each limit cycle. These entropies can be defined with the assistance of the Heaviside step function, $u(\mu)$, where $\mu$ is the parameter. Defining the period two limit cycle as beginning at $\mu_1$, the period four limit cycle beginning at $\mu_2$, etc., we can define the entropies as follows

\begin{eqnarray}
H_1&=&\log 2 u(\mu-\mu_1)\\
H_2&=&\log 2 u(\mu-\mu_1)+\log 2 u(\mu-\mu_2) \nonumber\\ 
H_3&=&\log 2 u(\mu-\mu_1)+\log 2 u(\mu-\mu_2)+ \nonumber\\
&&\log 2 u(\mu-\mu_3) \nonumber\\ 
&\dots & \nonumber
\end{eqnarray}

We can then see that we can define higher order entropies in terms of lower order entropies.

\begin{eqnarray}
H_2&=&H_1+\log 2 u(\mu-\mu_2)\\
H_3&=&H_2+\log 2 u(\mu-\mu_3) \nonumber\\
&\dots&\nonumber
\end{eqnarray}

Finally, we can easily define the conditional entropies.
\begin{eqnarray}
H(2)&=&H_2-H_1=\log 2 u(\mu-\mu_2)\\
H(3)&=&H_3-H_2=\log 2 u(\mu-\mu_3) \nonumber\\
&\dots& \nonumber
\end{eqnarray}

A key question is how to relate these entropies to the first Feigenbaum constant, $\delta$. One approximation can be used given the integral of the first-order entropy across the interval $[\mu_1,\mu_{\infty}]$. Using the definition of the integral of the Heaviside function as the step function,

\begin{equation}
\int_{\mu_1}^{\mu_{\infty}} H_1 d\mu = \log 2 \int_{\mu_1}^{\mu_{\infty}} u(\mu-\mu_1) d\mu
\end{equation}

and

\begin{equation}
\log 2 \int_{\mu_1}^{\mu_{\infty}} u(\mu-\mu_1) d\mu = \log 2 (\mu_{\infty} - \mu_1)
\end{equation}

\begin{table}[t!]
\centering
\large{
\begin{tabular}{|c|c|c|c|c|}
\hline
$\beta$&$\frac{2\varphi+\beta}{\log 2 + \beta}$&$\delta$, $\alpha$ act.&\% difference&Comment \\
\hline
0&4.66866&4.6692&0.01\%&$\delta$ for $n=2$\\
\hline
1&2.50189&2.50291&0.04\%&$\alpha$ for $n=2$\\
\hline
2&1.94422&1.92769&-0.86\%&$\alpha$ for $n=3$\\
\hline
3&1.68855&1.6903&0.1\%&$\alpha$ for $n=4$\\
\hline
4&1.54184&1.55577&0.9\%&$\alpha$ for $n=5$\\
\hline
5&1.44666&1.46774&1.44\%&$\alpha$ for $n=6$\\
\hline
\end{tabular}
}
\caption{Table of values of Feigenbaum's constants for function of degree $n$. Estimated values, actual values computed numerically \citep{constants}, percent differences, and explanations are given.}
\label{feigtable}
\end{table}

This expression can be combined with the well-known universal scaling approximation

\begin{equation}
\mu_{\infty} - \mu_n \approx \frac{C_1}{\delta^n}
\end{equation}

where $C_1$ is a constant. By setting $n=1$ and creating a new constant, $C_2$, that is defined by 

\begin{equation}
C_2 = \frac{C_1\log 2}{\mu_{\infty} - \mu_1}
\end{equation}

We can investigate the value of $C_2$ with respect to $\log 2$ and $\delta$ in

\begin{equation}
\delta \approx \frac{C_2}{\log 2}
\end{equation}

Here the results become interesting, though also very baffling. First, $C_2 = \delta \log 2$ or 3.236, which is approximately $2\varphi$, where $\varphi$ is the golden ratio or $\frac{1+\sqrt{5}}{2}$. So an estimate of $\delta$ can be reached by

\begin{equation}
\delta \approx \frac{2\varphi}{\log 2}
\label{cooleqn1}
\end{equation}

This estimate is accurate to within the 3rd decimal point or about 0.01\% (see Table \ref{feigtable}) of the correct value 4.669202\dots. By itself, this is unimpressive. While it may be slightly more elegant, in fact it is not even the most accurate estimate of $\delta$ compared to other expressions such as $\delta \approx \pi + \tan^{-1}(e^{\pi})$ which is good for six decimal places. However, one discovers that by changing equation \ref{cooleqn1} to sum identical positive integers ($\beta$) to the numerator and denominator of the form

\begin{equation}
\frac{2\varphi + \beta}{\log 2 + \beta}
\end{equation}

we can incredibly derive estimates of other Feigenbaum constants. For $\beta=1$, the result is 2.5019, only 0.04\% different from Feigenbaum's second constant, $\alpha$. Further, for $\beta=2$, the result is 1.944, a close (0.9\%) estimate for $\alpha$ for cubic functions whose value is 1.9277. This pattern continues for the $\alpha$ of higher degree functions as seen in Table \ref{feigtable}.

Obviously, all of these values are only close approximations and the approximations become increasingly inaccurate as $\beta$ increases. As $\beta$ goes to infinity, the value of $\alpha$ converges to 1.

\section{Conclusion}

In conclusion, we have introduced two main concepts: first the concept of using conditional entropies to define limit cycles and period doubling. Though this paper only handled period doubling you can also investigate period tripling, quadrupling, etc. in a similar manner using alphabets with the required number of items and using $\log 3$, $\log 4$, etc. and their multiples to define the different orders of entropy. Second, in attempting to combine this insight with the estimation of Feigenbaum's constant we have found an intriguing approximation for $\delta$ and $\alpha$ for multiple classes of functions using primarily $2\varphi$ and $\log 2$. These results, despite being only approximations, seem to show that not only are both of the constants related to $2\varphi$ and $\log 2$ but they are also related to each other in a fundamental way. Perhaps these relationships will help us exactly calculate the underlying expression which gives the Feigenbaum constants.

\end{document}